\documentclass[12pt]{article}
\usepackage{amsfonts}
\newcommand{\Fd}{\widetilde{F}}
\newcommand{\n}{\eta}
\newcommand{\w}{\omega}
\begin{document}
%%%%%%%%%%%%%%%%%%%%%%%%%%%%%%%%%%%%%%%%%%%%%%%%%%%%%%%%%%%%%%%%%%%%%
\title{Fermionic determinant with a linear domain wall in $2+1$
  dimensions} 
\author{L.~Da~Rold~$^a$~\footnote{Electronic address:
    daroldl@ib.cnea.gov.ar}, C.~D.~Fosco~$^a$~\footnote{Electronic
    address: fosco@cab.cnea.gov.ar} and
  A.~P.~C.~Malbouisson$^b$~\footnote{Electronic address:
    adolfo@cbpf.br}
  \\
  \\
  \normalsize{\it $^a$Centro At\' omico Bariloche - Instituto
    Balseiro,}\\ \normalsize{\it Comisi{\'o}n Nacional de Energ\'{\i}a
    At{\'o}mica}\\ \normalsize{\it 8400 Bariloche, Argentina.}\\
  \\
  \normalsize{\it $^b$Centro Brasileiro de Pesquisas F\'{{\i}}sicas - CBPF/MCT,}\\
  \normalsize{\it Rua Dr. Xavier Sigaud, 150}\\
  \normalsize{\it 22290-180 Rio de Janeiro RJ, Brasil.}}
\date{\today} \maketitle
%====================================================================
\begin{abstract}
\noindent We consider a Dirac field in $2+1$ Euclidean dimensions, 
in the presence of a linear domain wall defect in its mass, and a
constant electromagnetic field.  We evaluate the exact fermionic
determinant for the situation where the defect is assumed to be
rectilinear, static, and the gauge field is minimally coupled to the
fermions. We discuss the dependence of the result on the (unique)
independent geometrical parameter of this system, namely, the relative
orientation of the wall and the direction of the external field.  We
apply the result for the determinant to the evaluation of the vacuum
energy.
\end{abstract}
\newpage
\section{Introduction}
In general, the fermionic action $S$ is (or may be transformed into) a
quadratic expression in the Grassmann fields, which is in turn
determined by a Dirac operator ${\mathcal D}$. This Dirac operator,
except for some trivial cases, carries a dependence on some parameters
and fields. At some stage in the resolution of a dynamical problem, it
may be useful to regard those fields as `external', either because
they don't have a proper dynamics, or because they have not yet been
quantized.

The fermionic determinant is formally given by the product of the
eigenvalues of the Dirac operator. The explicit calculation of those
eigenvalues is, however, impossible to achieve for an arbitrarily
general situation, and one has to resort to some form of approximate
expansion.  Nevertheless, there are many highly symmetrical external
field configurations where this explicit diagonalization can be
successfully carried out.  In particular, the constant $F_{\mu\nu}$ case
in several spacetime dimensions, allows for the determination of the
exact eigenvalues, since the diagonalization of ${\mathcal D}$ reduces
in this case to finding the spectrum of a one dimensional harmonic
oscillator~\cite{schw}. Many interesting results have been found in
this area, and they have application, for example, to the
determination of effective actions~\cite{eff}.

In most of these cases, the method of resolution amounts to expressing
first the Dirac operator ${\mathcal D}$ as a quadratic form in the
operators $x_\mu$, $p_\mu$~\footnote{We work with Euclidean coordinates:
  $x_\mu=(x_0, x_1, x_2)$, where $x_0$ denotes the Euclidean time.} (as
in the Fock-Schwinger method~\cite{schw}).  The diagonalization
procedure is then equivalent to defining a transformation from $x_\mu$
and $p_\mu$ to a new set of canonical operators, and thus may in general
be expressed as a Bogoliubov transformation in some suitable creation
and annihilation operators.  The original problem becomes then an
algebraic one, with a complexity which depends of course on the kind
of situation considered.

In this article we will evaluate the fermionic determinant
corresponding to a Dirac field in $2+1$ dimensions, coupled to an
external uniform electromagnetic field $F_{\mu\nu}$, and with a mass term
which is a linear functional of the coordinates. This kind of mass
term is a simple example of a situation where a domain wall like
defect (of constant slope) is present, in this case in the
parity-breaking mass term. We note that, if the electromagnetic field
were parallel to the defect, the situation would fall into the
well-known Callan-Harvey phenomenon~\cite{callan}, by which a chiral
fermionic zero mode is induced on the defect. We will, however,
consider situations where the field has a different orientation.

We will show that the modulus of the determinant can be found exactly
for these configurations, what represents a non-trivial generalization
of the well-known situation corresponding to a uniform electromagnetic
field and a {\em constant\/} mass term.

The Euclidean action $S$ for this system is
\begin{equation}\label{accion}
S\;=\; \int d^3x \,\bar{\Psi} {\mathcal D} \Psi \;,
\end{equation}
where ${\mathcal D}=\not\!\!\!\partial+i\not\!\!A+M$, and we have absorbed
the electric charge into the definition of the gauge field $A$. The
Hermitian $\gamma$-matrices are in the irreducible $2\times 2$ representation of
the Dirac algebra:
\begin{equation}\label{gammas}
\{\gamma_{\mu},\gamma_{\nu}\}\,=\,2\delta_{\mu\nu} \;.
\end{equation}

The domain wall, defined as the region where the mass changes sign,
will in our case have only two defining properties: its location and
its slope.  The latter determines the localization (or not) of the
defect, and is quantitatively measured by the normal derivative of the
mass along the curve of the defect. The situation we consider in this
paper may be considered as an approximate treatment of localized
defects; a general discussion on the localization of modes for this
system may be found in~\cite{paper2+1}.

The Dirac operator ${\mathcal D}$ appearing in (\ref{accion}) is not
the more suitable one to deal with the eigenvalue problem, since it is
not Hermitian.  It is convenient to consider, instead, a related
operator ${\mathcal H}$ which is always Hermitian, has a complete set
of eigenstates, and its eigenvalues are the squares of the Dirac
operator's eigenvalues.  This operator ${\mathcal H}$ is of course
\begin{equation}\label{hermitico}
{\mathcal H}={\mathcal D}^\dagger {\mathcal D}\;,
\end{equation}
which is indeed the operator defining the positive Hermitian part of
${\mathcal D}$ in its polar decomposition.  We can see from the
structure of ${\mathcal H}$, that for a mass linear in the
coordinates, and with a constant electromagnetic potential, it will be
a quadratic expression in $x_\mu$ and $p_\mu$.  The mass will be regarded
as a linear functional of the coordinates.

This kind of configuration for the mass and the electromagnetic field
can be characterized by two vectors in Euclidean spacetime. For the
mass, vanishing along a line that passes through the origin (what is
always possible by a proper choice of coordinates), can be defined in
terms of a vector $\n_\mu$,
\begin{equation}\label{M}
M\;=\;\n_\mu x_\mu \;,
\end{equation}
so that $\n_\mu$ points in the direction normal to the plane that
defines the defect hyperplane.  Regarding the electromagnetic
potential $A_\mu$, a linear function of the coordinates suffices to
generate a constant electromagnetic field $F_{\mu\nu}$, since the gauge
potential can be written in a symmetric gauge as follows:
\begin{equation}\label{A}
A_\mu\;=\;-\frac{1}{2}F_{\mu\nu}x_\nu \;.
\end{equation}

We have found it convenient to use, rather than the antisymmetric
tensor, $F_{\mu\nu}$, its dual $\Fd_\mu$,
\begin{equation}\label{Fd}
\Fd_\mu\;=\;\frac{1}{2}\epsilon_{\mu\nu\rho}F_{\nu\rho},
\end{equation}
where $\epsilon_{\mu\nu\rho}$ is the totally antisymmetric tensor in $3$ dimensions.
This puts both the defect and the gauge field on an equal footing,
simplifying the study of the dependence of the fermionic determinant
on the geometric invariants that may be built out of those external
fields.

Writing $A_\mu$ in terms of the dual of $F_{\mu\nu}$, $A_\mu=\frac{1}{2}
\epsilon_{\mu\nu\rho} \Fd_\rho x_\nu$, one can present the Dirac operator parametrized by
two constant vectors, thus summarizing all its dependence on the
external parameters:
\begin{equation}\label{D}
{\mathcal D}\;=\;\gamma \cdot \partial - \frac{i}{2} \gamma \cdot x \times \Fd + \n \cdot x \;,
\end{equation}
where we have used the notation
\begin{equation}\label{productos}
(a \times b)_\mu \;\equiv\; \epsilon_{\mu\nu\rho} a_\nu b_\rho\; ,
\;\;\;\;\;\;  a \cdot b\;\equiv\;a_\mu b_\mu \;.
\end{equation}

The relative orientation between $\n_\mu$ and $\Fd_\mu$ covers all the
freedom to describe the different geometrical configurations allowed
within this model for the Dirac operator. Different particular cases
will describe rather different situations, both from the physical and
the mathematical points of view. The simplest situation corresponds to
$\eta$ parallel to ${\tilde F}$, as in the Callan and Harvey mechanism,
since the gauge field is then entirely contained in the defect
worldsheet. Then the determinant of the three dimensional system
immediately factorizes into the product of an infinite number of
(decoupled) determinants in $2+1$ dimensions~\cite{paper2+1}.  Our
presentation begins, in section~\ref{sec:first}, with the more
interesting case $\Fd \cdot \n=0$.  Namely, the electromagnetic field is
normal to the defect hypersurface.  The interest in this case stems
from the fact that it takes into account the competition between the
localizing effect due to the defect and the effect of the
electromagnetic field, which will tend to `drag' any charge
distribution in a direction normal to the defect.  An application to
the calculation of the vacuum energy, for this case, is presented in
section~\ref{sec:vacuum}.  Finally, we conclude by deriving the
eigenvalues of the equivalent diagonal operator for the general case
in section~\ref{sec:general}.

\section{Fermionic determinant for $\Fd \cdot \n=0$}\label{sec:first}
In this situation, $\Fd_\mu$ and $\n_\mu$ define two orthogonal directions
under the scalar Euclidean product defined in (\ref{productos}).
Thus, in terms of these two vector fields, we may construct two
families of curves in the Euclidean 3-dimensional space that may be
used to define coordinates, varying along their integral lines.  A
third family of curves, orthogonal to the two previous ones, can then
be uniquely defined by selecting a right-handed orientation.  We
define $\hat{t}_\mu$ to be the unit vector in the direction of $\Fd_\mu$.
It is obvious from equation (\ref{D}) that in the particular case we
are dealing with, the operator ${\mathcal D}$ does not depend on the
coordinate $x_t$ (which parametrizes the integral curves of
$\hat{t}_\mu$).  We see that it is then sufficient to diagonalize the
part of ${\mathcal H}$ which depends on the other two coordinates.
Indeed, using equation (\ref{D}), we can express ${\mathcal H}$ in
terms of the external parameters,
\begin{equation}\label{Htotal}
{\mathcal H}\;=\;-\partial^2\,+\,\frac{1}{4}[x^2\Fd^2-(x \cdot \Fd)^2]+(x \cdot \n)^2
\,-\,i (x \times \partial) \cdot \Fd\,+\,\gamma \cdot (\Fd-\n)\;,
\end{equation}
which, in the coordinates defined by the unit vectors
\begin{equation}
\hat{\n} \equiv \frac{\n}{\|\n\|},\;\;\;\; \hat{t} \equiv
\frac{\Fd}{\|\Fd\|},\;\;\;\; \hat{b} \equiv \hat{\n}\land
\hat{t}\;,
\end{equation}
becomes,
\begin{equation}\label{Hcoord}
{\mathcal H}\;=\;-\partial_\n^2-\partial_t^2-\partial_b^2+(\frac{\Fd^2}{4}+\n^2)x_\n^2
+\frac{\Fd^2}{4}x_b^2-i\Fd (x_b\partial_\n-x_\n\partial_b)+\gamma \cdot (\Fd-\n)\;.
\end{equation}
In this expression, ${\mathcal H}$ is invariant under translations in
$x_t$, thus the diagonalization of ${\mathcal H}$ amounts to
solving the reduced problem defined by a different Hamiltonian $H$
acting on functions depending on the two coordinates $x_\n$ and $x_b$,
\begin{equation}
{\mathcal H} \;\equiv\; -\partial_t^2\,+\,2H, 
\end{equation}
where
\begin{equation}\label{H}
H\;=\;\frac{1}{2}p_\n^2+\frac{1}{2}p_b^2+\frac{\omega_n^2}{2}x_\n^2
+\frac{\omega_b^2}{2}x_b^2 + \frac{\Fd}{2} (x_b p_\n-x_\n p_b)
+\frac{1}{2} \gamma \cdot (\Fd-\n) \;.
\end{equation}
In (\ref{H}) we have introduced the constants
\begin{equation}\label{w}
\w_\n^2=\frac{\Fd^2}{4}+\n^2,\;\;\;\;\;\;\;\w_b^2=\frac{\Fd^2}{4}\;.
\end{equation}
To study the diagonalization of $H$, it is first convenient to define
suitable creation and annihilation operators for each coordinate,
\begin{eqnarray}\label{a}
a_j&=&\frac{1}{\sqrt{2}}(\w_j^{1/2}x_j+i\w_j^{-1/2}p_j)
\nonumber,\\
a_j^\dagger&=&\frac{1}{\sqrt{2}}(\w_j^{1/2}x_j-i\w_j^{-1/2}p_j),\;\;\;\;
j=\n,b
\end{eqnarray}
(no sum over $j$), since the expression for $H$ becomes more
symmetric, and its diagonalization can be done by a generalization of
the usual procedure for a quadratic Hamiltonian.
  
In terms of these operators, $H$ contains both bilinear and constant
terms:
$$
H \;=\; \w_\n a_\n^\dagger a_\n + \w_b a_b^\dagger a_b +
\frac{\|\Fd\|}{4i}(\sqrt{\frac{\w_\n}{\w_b}}-\sqrt{\frac{\w_b}{\w_\n}})
(a_\n a_b - a_\n^\dagger a_b^\dagger)
$$
$$
+\;\frac{\|\Fd\|}
{4i}(\sqrt{\frac{\w_\n}{\w_b}}+\sqrt{\frac{\w_b}{\w_\n}}) (a_\n a_b^\dagger
- a_\n^\dagger a_b)
$$
\begin{equation}
\label{Ha}
+ \frac{1}{2}[\w_\n + \w_b + \gamma \cdot (\Fd-\n)]\;,
\end{equation}
which has the important property of containing terms which are
quadratic in the creation and annihilation operators, and not just the
standard one with creation and annihilation operators mixed by a
Hermitian matrix. This makes the diagonalization more cumbersome.
Indeed, to diagonalize $H$, we first introduce some $2\times2$ matrices,
defined by
\begin{equation}\label{bilinear}
H\;=\;a_j^\dagger A_{jl} a_l + \frac{i}{2} a_j B_{jl} a_l -
\frac{i}{2} a_j^\dagger B_{jl} a_l^\dagger +  \frac{1}{2}[\w_\n +
\w_b + \gamma \cdot (\Fd-\n)]\;,
\end{equation}
where $A$ is Hermitian while $B$ is real and symmetric
\begin{equation}\label{matrices}
A= \left(
\begin{array}{cc}
  \w_\n & -iu \\
  iu   & \w_b
\end{array}
\right) \;\;\;\;\; B= \left(
\begin{array}{cc}
  0 & t \\
  t & 0
\end{array}
\right),
\end{equation}
with the constants $u$ and $t$ defined by
\begin{equation}\label{ut}
u=\frac{\|\Fd\|}{4i}(\sqrt{\frac{\w_\n}{\w_b}}+\sqrt{\frac{\w_b}{\w_\n}}),\;\;\;\;
t=\frac{\|\Fd\|}{4i}(\sqrt{\frac{\w_\n}{\w_b}}-\sqrt{\frac{\w_b}{\w_\n}})\;.
\end{equation}
The diagonalization is non trivial due to the presence of the $B$
matrix, which only vanishes for $\eta=0$.  To diagonalize the bilinear
term in equation (\ref{bilinear}), we shall need to define new
creation and annihilation operators, which will be linear combinations
of the ones defined in Eq.~\ref{a}. We first note that the $a_j,
a_j^\dagger$ operators have been defined in such a way that they verify the
commutation relations:
\begin{eqnarray}\label{aconmutacion}
& &[a_j^\dagger,a_l^\dagger]=[a_j,a_l]=0, \nonumber\\ &
&[a_j,a_l^\dagger]=\delta_{jl}, \nonumber\\ &
&[H,a_l^\dagger]=a_j^\dagger A_{jl} + i a_j B_{jl}, \nonumber\\ &
&[H,a_l]=-a_j A^*_{jl} + i a_j^\dagger B_{jl},
\end{eqnarray}
where $A^*$ is the complex conjugate of $A$. We require the new
operators $b_j, b_j^\dagger$ to be also canonical and to diagonalize the
bilinear form of (\ref{bilinear}). Following~\cite{creation}, we
require them to satisfy the commutation relations,
\begin{eqnarray}\label{bconmutacion}
& &[b_j^\dagger,b_l^\dagger]=[b_j,b_l]=0,\nonumber\\ &
&[b_j,b_l^\dagger]=\delta_{jl}, \nonumber\\ &
&[H,b_l^\dagger]=b_l^\dagger E_l,\nonumber\\ & &[H,b_l]=-b_l E_l
\end{eqnarray}
(no sum over $l$). We can write the new operators in terms of the 
old ones by using a general Bogoliubov transformation of the following
kind: 
\begin{eqnarray}\label{cambiobase}
b_l^\dagger=a_j^\dagger U_{jl} - a_j V_{jl},\nonumber \\
b_l=-a_j^\dagger V^*_{jl} + a_j U^*_{jl},
\end{eqnarray}
where, to fulfill the commutation relations above, the coefficients
$U$ and $V$ must verify the matrix equations
\begin{equation}\label{autovalores0}
U^\dagger U - V^\dagger V\,=\,I  \;\;,
\end{equation}
\begin{equation}\label{autovalores1}
U^t V-V^t U\,=\,0 \;\;,
\end{equation}
\begin{equation}\label{autovalores2}
A U-i B V\,=\, U E \;\;,
\end{equation}
and
\begin{equation}\label{autovalores3}
i B U + A^* V \;=\;- V E \;\;,
\end{equation}
$E$ being the diagonal matrix whose elements are the eigenvalues of
$H$ and $U^{\dagger}$, as usual, stands for the adjoint of $U$.  Similar
redefinitions, although for a different case, where used 
in~\cite{creation} and~\cite{bogolubov}.  

The system of equations involving the coefficient matrices defines a
generalized diagonalization problem, which involves matrices rather
than vectors. The next step is to obtain $E$ explicitly in terms of
the parameters of the theory.  We first note that not all the
equations that define this eigensystem are independent. To determine
$E$ we may, for example, use Equation~(\ref{autovalores2}) to express
the coefficients $V_{jl}$ in terms of $U_{jl}$, the parameters of the
theory and the eigenvalues $E_l$. This yields
\begin{equation}\label{Vjl}
V\;=\; \left(
\begin{array}{cc}
r_{11}U_{11}+s_{11}U_{21} & r_{12}U_{12}+s_{12}U_{22}\\
r_{21}U_{21}+s_{21}U_{11} & r_{22}U_{22}+s_{22}U_{12}
\end{array}
\right),
\end{equation}
where the coefficients $r_{jl}$ and $s_{jl}$ are defined by
\begin{equation}
r\;=\; \left(
\begin{array}{cc}
  \frac{u}{t} & \frac{u}{t} \\
  -\frac{u}{t} & -\frac{u}{t}
\end{array}
\right), \;\;\;\;\;\; s= \left(
\begin{array}{cc}
  \frac{i(E_1-\w_b)}{t} & \frac{i(E_2-\w_b)}{t} \\
  \frac{i(E_1-\w_\n)}{t} & \frac{i(E_2-\w_\n)}{t}
\end{array}
\right).
\end{equation}
Introducing this result into Eq.~(\ref{autovalores3}) yields two
equations, now involving $U_{11}$ and $U_{21}$, plus two others
involving $U_{12}$ and $U_{22}$,
\begin{eqnarray}\label{U11}
U_{21} &=& U_{11} \frac{2 i \w_\n u}{E_1^2+t^2-E_1 \w_b-u^2+E_1\w_\n-\w_\n \w_b} \nonumber\\
&=&i U_{11} \frac{E_1^2+t^2-E_1 \w_\n-u^2+E_1\w_b-\w_\n \w_b}{2\w_b u},
\end{eqnarray}
and
\begin{eqnarray}\label{U22}
U_{12}&=&-U_{22}\frac{2 i \w_b u}{E_2^2+t^2-E_2 \w_\n-u^2+E_2\w_b-\w_\n \w_b} \nonumber\\
&=&- i U_{22} \frac{E_2^2+t^2-E_2 \w_b-u^2+E_2\w_\n-\w_\n \w_b}{2\w_\n u}.
\end{eqnarray}

Demanding equations (\ref{U11}) and (\ref{U22}) to be consistent,
implies a set of constraints for the eigenvalues $E_l$. After some
algebra, those constraints may be written in terms of just one
equation, that determines the possible values of $E_l$ in terms of the external
parameters. There are $4$ solutions to this equation: besides the
double eigenvalue $E_1=0$, we find:
  \begin{equation}
  E_2\,=\,-E_3\,=\,\sqrt{\w_\n^2+3\w_b^2}\;.
\label{eigenvalues1}
\end{equation}
This set of solutions contains the true eigenvalues; however, to
discard the spurious ones, we should check whether they are
consistent with (\ref{autovalores0}) and (\ref{autovalores1}), which
guarantee the canonical commutation relations for the `new' operators.
In terms of the variables defined in Eq.(\ref{Vjl}),
(\ref{autovalores1}) can be recast in the form,
\begin{equation}\label{A=0}
pq(s_{22}-s_{11})+q(r_{22}-r_{21})+p(r_{12}-r_{11})+(s_{12}-s_{21})=0,
\end{equation}
where $p$ and $q$ are defined by:
\begin{equation}\label{p}
p\;=\;-i\frac{E_j^2+t^2-E_j \w_b-u^2+E_j \w_\n-\w_\n \w_b}{2 \w_\n u},
\end{equation}
and
\begin{equation}\label{q}
q\;=\;i\frac{2 \w_\n u}{E_k^2+t^2-E_k \w_b-u^2+E_k \w_\n-\w_\n \w_b}
\end{equation}
where the labels $j$ and $k$ stand for the two different eigenvalues
that can be chosen from the four possibilities in
Eq.(\ref{eigenvalues1}).

Eq.(\ref{autovalores0}) splits into three algebraic equations, one
corresponding to the non-diagonal elements,
\begin{equation}\label{C=0}
p+q^*-(r_{11}+q s_{11})^*(r_{12}p+s_{12})-(r_{21}q+
S_{21})^*(r_{22}+p s_{22})=0,
\end{equation}
plus two others for the diagonal terms,
\begin{equation}\label{B=1} 
|U_{11}|^2[1+|q|^2-|r_{11}+q s_{11}|^2-|r_{21}q+s_{21}|^2]\;=\;1,
\end{equation}
\begin{equation}\label{D=1}
|U_{22}|^2[1+|p|^2-|r_{12}p+ s_{12}|^2-|r_{22}+p s_{22}|^2]\;=\;1\;.
\end{equation}
Eqs.~(\ref{A=0}) and (\ref{C=0}) do not contain the coefficients of
the matrices $U$ and $V$, and thus they are equations for the
eigenvalues $E_l$.  In terms of the original parameters, we obtain for
the eigenvalues of $H$,
\begin{equation}
E_1=0,\;\;\;\; E_2=\sqrt{\w_\n^2+3\w_b^2}=\sqrt{\Fd^2+\n^2}\;.
\end{equation}
Note that (\ref{B=1}) and (\ref{D=1}) only fix the moduli of $U_{11}$
and $U_{22}$; the phase of these coefficients is not fixed by the
basis choice, and on the other hand no physical magnitude will depend
on it.

We note that one of the eigenvalues vanishes, what implies
$|U_{11}|\to\infty$. This means that, in the direction corresponding to the
operators $b_1$ and $b_1^\dagger$, there is no harmonic mode, but rather a
free motion. Thus we define the corresponding conjugate $x_1$ and
$p_1$ variables, such that ${\mathcal H}$ in (\ref{Hcoord}) becomes
\begin{equation}\label{Hdiag}
{\mathcal H}\;=\; p_t^2 + p_1^2 + 2 E_2 b_2^\dagger b_2 + \w_\n + \w_b + \gamma \cdot (\Fd-\n)\;.
\end{equation}
We have thus obtained the diagonal form for the operator
$\mathcal{H}$.  From this expression, we can obtain the modulus of the
determinant of the Dirac operator ${\mathcal D}$.
\section{Vacuum energy in a constant electromagnetic 
  field}\label{sec:vacuum}
In this section we evaluate the vacuum energy for the fermionic
system studied in the previous section (i.e., when $\Fd \cdot \n =0$); we shall 
see that this physical magnitude is entirely determined by
(\ref{Htotal}). 

To that end we first consider the vacuum energy density for a
fermionic system in the presence of an external electromagnetic field.
The vacuum to vacuum transition probability amplitude is given by the
$\mathcal S$ matrix expectation value between vacuum states, which depends
on the external potential $A$,
\begin{equation}
{\mathcal S}_0(A)=< 0 \;in | {\mathcal S} | 0 \;in > \;.
\end{equation}
This object is usually normalized with respect to the transition
amplitude in the absence of external fields, ${\mathcal S}_0(A)$:
\begin{equation}\label{s0}
|{\mathcal S}'_0(A)|^2 \equiv \frac{|{\mathcal S}_0(A)|^2}
{|{\mathcal S}_0(0)|^2}\;.
\end{equation}
In terms of the normalized transition amplitude ${\mathcal S}'_0(A)$,
one then defines a local function $w(x)$~\cite{itzykson}
\begin{equation}\label{w}
|{\mathcal S}'_0(A)|^2 \equiv exp[-\int d^nx\;w(x)] \;,
\end{equation}
where $n$ denotes the spacetime dimension. In the Euclidean
formulation we can interpret $w(x)$ as half the vacuum energy density.
To see this we consider the Euclidean evolution operator in the
interaction representation $U(\beta,\beta_0)$, which obeys the
equations~\cite{zinnjustin},
\begin{equation}\label{S0euclideo}
\lim_{\beta\to\infty}[Tr\,U(\frac{\beta}{2},-\frac{\beta}{2})]=\langle0, in|U(\infty,-\infty)|0,in \rangle = S'_0(A)
\end{equation}
and
\begin{equation}\label{E0euclideo}
\lim_{\beta\to\infty}[-\frac{1}{\beta}\ln Tr\,U(\frac{\beta}{2},-\frac{\beta}{2})]=E_0,
\end{equation}
where $E_0$ is the vacuum energy. From these equations and the
definition (\ref{w}) we see that the integral of $w(x)$ over the
spatial coordinates is half the vacuum energy. In order to obtain
$w(x)$, we  first have to evaluate ${\mathcal S}_0(A)$. In the functional
integral representation, we may write
\begin{equation}\label{svacio}
{\mathcal S}_0(A)=|{\mathcal N}|^2 \int D\bar{\Psi} D\Psi e^{-\int d^3x \bar{\Psi} {\mathcal D}[A] \Psi},
\end{equation}
where the notation ${\mathcal D}[A]$ indicates the Dirac operator
dependence on the external field $A$. Since the integral is over
Grassmannian variables, we write
\begin{equation}\label{det}
{\mathcal S}_0(A)=|{\mathcal N}|^2 \det({\mathcal D}[A]),
\end{equation}
where $\det$ stands for the determinant over both spinor and functional
spaces. Inserting (\ref{det}) into the definition (\ref{s0}), we see that
\begin{equation}\label{sdet}
|{\mathcal S}'_0(A)|^2=exp\{Tr\;\ln{\mathcal H}[A]-Tr\;\ln{\mathcal H}[0]\}.
\end{equation}

Now,  we use the Frullani's identity:
\begin{equation}
\ln \frac{a}{b}=\lim_{\epsilon \to 0} \int_\epsilon^\infty \frac{ds}{s}(e^{-sb}-e^{-sa})
\end{equation}
to rewrite equation (\ref{sdet}) in a particularly convenient integral
representation:
\begin{equation}\label{str}
\log(|{\mathcal S}'_0(A)|^2)= Tr \lim_{\epsilon \to 0} \int_\epsilon ^\infty \frac{ds}{s} (e^{-s{\mathcal H}[A]}-e^{-s{\mathcal H}[0]}),
\end{equation}
whence we can obtain the vacuum energy density as: 
\begin{equation}\label{wtr}
w(x) = tr \lim_{\epsilon \to 0} \int_\epsilon ^\infty \frac{ds}{s} (e^{-s{\mathcal H}[A]}-e^{-s{\mathcal H}[0]}),
\end{equation}
where now the trace only affects the spinor space indices.

We evaluate now the vacuum energy, the integral of $w(x)$ over all the
Euclidean space, for the previously described case. The operators
${\mathcal H}[A]$ and ${\mathcal H}[0]$ are respectively given by,
\begin{equation}\label{HA}
{\mathcal H}[A]=p_t^2 + p_1^2 + 2b_2^\dagger b_2 E_2 + \w_n + \w_b +\gamma \cdot (\Fd-\n)
\end{equation}
and
\begin{equation}\label{H0}
{\mathcal H}[0]= p^2 + M^2 - \gamma \cdot \n \;.
\end{equation}
We can write ${\mathcal H}[0]$ in terms of some creation and
annihilation operators $c$ and $c^\dagger$, defined as in equation
(\ref{a}), with frequency $\w=\|\n\|$, as follows:
\begin{equation}\label{H0c}
{\mathcal H}[0]= p_t^2 + p_b^2 + \|\n\|(2c^\dagger c+1) - \gamma \cdot\n \;.
\end{equation}
Evaluating the trace, we see that:
$$ w(x)\,=\, 2 \,\lim_{\epsilon \to 0} \int_\epsilon ^\infty\frac{ds}{s} \left\{ \langle x|e^{-s[p_t^2+p_1^2+2b_2^\dagger
b_2+\w_\n+\w_b]}(e^{sE_2}-e^{-sE_2}) |x\rangle \right.$$ 
\begin{equation}\label{wtr1}
\left. -\, e^{-s[p_t^2+p_b^2+2c^\dagger c+\|\n\|]}(e^{s\|\n\|}-e^{-s\|\n\|})\}|x\rangle\right\} \;.
\end{equation}

The system we are considering is not translation invariant, and
moreover, we may write the explicit coordinate dependence of $w(x)$ by
using the completeness of the eigenstates of the operators $b_2^\dagger b_2$
and $c^\dagger c$ for the respective Hilbert spaces. Denoting by
$\langle x_2|m_{2}\rangle$ and $\langle x_\eta|m_{\n}\rangle$ the respective eigenstates of the
operators $b_{2}^{\dagger}b_{2}$ and $c^{\dagger}c$, which are as usual labeled by
a non-negative integer $m=0,1,\ldots,\infty$, then we may write $w(x)$ as
$$
w(x)\;=\; 2 \lim_{\epsilon \to 0} \int_\epsilon^\infty \frac{ds}{s} \sum_{m=0}^\infty \left\{ |\langle x_2|m_2\rangle|^2 \,
\langle x_{\n} x_b|e^{-s[p_t^2+p_1^2+2m+\w_\n+\w_b]}|x_{\n} x_b\rangle 
\right.$$
\begin{equation}\label{wtr2}
\left. (e^{sE_2}-e^{-sE_2}) \,-\, |\langle x_{\eta}|m_\n \rangle|^2 
\langle x_t x_b|e^{-s[p_t^2+p_b^2+2m+\|\n\|]}|x_t x_b\rangle 
(e^{s\|\n\|}-e^{-s\|\n\|}) \right\}\;.
\end{equation}
It should be obvious then that each term in $w(x)$ in in fact a 
function of only one coordinate, $x_2$. Indeed, one may even check that
each term in the series above may be regarded as a $1+1$ dimensional
determinant times the square of the amplitude for the corresponding
harmonic oscillator mode.

The vacuum energy density in terms of the eigenstates of harmonic
modes $b_{2}^{\dagger}b_{2}$ and $c^{\dagger}c$. To obtain the total vacuum
energy, $W$, we have to integrate over the phase space. After this
step the sum over $m$ can be performed and we obtain,
\begin{equation}\label{wtr2bis}
W\;=\;\frac{L^2}{2\pi}\,\lim_{\epsilon \to 0}w_{\epsilon}\equiv \frac{L^2}{2\pi} \lim_{\epsilon \to 0} \int_\epsilon
^\infty \frac{ds}{s^2}
[e^{-s(\w_\n+\w_b-E_2)}\coth(sE_2)-\coth(s\|\n\|)].
\end{equation}

The $L^2$ factor appears because we have considered the space volume
to be a square box. This equation expresses the vacuum energy of a
fermionic system with a domain wall that can be approximated linearly
near the defect, and in the presence of a constant electromagnetic
field, for a particular configuration ($\Fd \cdot \n=0$). We can see that
in the case where the electromagnetic field is absent the vacuum
energy vanishes, which is consistent with our non interacting fermions
normalization condition.

The regularized expression for $W$ diverges as $\epsilon$ goes to zero.  The
divergences in Eq.(\ref{wtr2bis}) should be isolated in some terms as
a first step and afterwards suppressed by some renormalization
procedure. To determine the divergent parts in $\epsilon$, we use the
expansion,
\begin{equation}
\coth(u)=\frac{1}{u}+\sum_{k=1}^{\infty}\frac{2^{2k}B_{2k}}{(2k)!}u^{2k-1}, 
\label{coth1}
\end{equation}
valid for $u^{2}<\pi^{2}$, $B_{2k}$, being the Bernoulli numbers.  We
obtain,
\begin{eqnarray}\label{sing1}
W_{\epsilon}&=& \frac{L^2}{2\pi}\{\frac{A^{2}}{E_{2}}\left[\Gamma(-2,\epsilon A)-\Gamma(-2,\frac{\pi A}{E_{2}})\right]
+B_{2}E_{2}\left[\Gamma(0,\epsilon A)-\Gamma(0,\frac{\pi  A}{E_{2}})\right] \nonumber\\
& + &  \sum_{k=2}^{\infty}\frac{2^{2k}B_{2k}E_{2}^{2k-1}A^{2-2k}}{(2k)!}
\left[\Gamma(2k-2,\epsilon A)-\Gamma(2k-2,\frac{\pi A}{E_{2}})\right]-\frac{|\n|}{2\pi^{2}} \nonumber\\ 
&+& \frac{1}{2|\n|\epsilon^{2}}-\log \left[ \frac{\pi}{|\n|}\right]^{-B_{2}|\n|}+B_{2}|\n|\log \epsilon
-\sum_{k=2}^{\infty}\frac{2^{2k}B_{2k}\pi^{2k-2}|\n|}{(2k)!} \nonumber\\
& +& F_{1}(w_{\n},w_{b},E_{2})+F_{2}(|\n|)\}\;, 
\end{eqnarray}
where
\begin{equation}
F_{1}(w_{\n},w_{b},E_{2})=\int_\frac{\pi}{E_{2}}^\infty dss^{-2}e^{-sA}\coth(sE_{2}) , \;  
F_{2}(|\n|)=\int_\frac{\pi}{|\n|}^\infty ds s^{-2}\coth(s|\n|)\;,
\label{F1F2}
\end{equation}
$A= w_{\n}+w_{b}-E_{2}$ and $\Gamma(\alpha,x)$ is the incomplete Gamma-function.
In the expression (\ref{sing1}) singularities are still present in the
terms $\Gamma(-2,\epsilon A)$ and $\Gamma(0,\epsilon A)$. To isolate them we use the formula,
\begin{equation}
\Gamma(\alpha,x)=e^{-x}x^{\alpha}\sum_{n=0}^{\infty}\frac{L_{n}^{\alpha}}{n+1},
\label{laguerre}
\end{equation}
where $L_{n}^{\alpha}$ are the Laguerre polynomials. We get at the limit $\epsilon
\to 0$, $\Gamma(-2,\epsilon A)\to \frac{1}{2A^{2}\epsilon^{2}}$ and $\Gamma(0,\epsilon A)\to \sum_{0}^{\infty}
\frac{1}{n+1}$. This last series diverges logarithmically.
After some rearrangements, we obtain,
\begin{eqnarray}
\label{sing2}
W_{\epsilon}&=& \frac{L^2}{2\pi}\{(\frac{1}{2E_{2}}-\frac{1}{2|\n|})\frac{1}{\epsilon^{2}}+ 
(B_{2}E_{2}\sum_{n=0}^{\infty}\frac{1}{n+1}+B_{2}|\n|\log \epsilon) \nonumber\\
&-& \frac{A^{2}}{E_{2}}\Gamma(-2,\frac{\pi A}{E_{2}})-B_{2}E_{2}\Gamma(0,\frac{\pi A}{E_{2}})- 
\log(\frac{\pi}{|\n|})^{-B_{2}|\n|} \nonumber \\ 
&+& \frac{A^{2}}{E_{2}}\sum_{k=2}^{\infty}\frac{B_{2k}}{(2k)!}(\frac{2E_{2}}{A})^{2k}
\left[\Gamma(2k-2,\epsilon A)-\Gamma(2k-2, \frac{\pi A}{E_{2}})\right] \nonumber \\
&-& \frac{|\n|}{\pi^{2}}(-\frac{1}{2}+\sum_{k=2}^{\infty}
\frac{(2\pi)^{2k}B_{2k}}{(2k)!})+F_{1}(w_{\n},w_{b},E_{2})+F_{2}(|\n|)\}.\nonumber \\
\end{eqnarray}

The two first terms in (\ref{sing2}), contain quadratic and
logarithmic divergencies which may be suppressed by a renormalization
procedure. It may be, at first sight, shocking to see that the
divergent part of $W_\epsilon$ is in fact not a finite degree polynomial in
the external field $F_{\mu\nu}$ and its derivatives, as the usual
divergence theorems for a standard Quantum Field Theory
imply~\cite{itzykson}. The reason for this seemingly contradictory
result is that the hypothesis for those theorems do not hold in the
present case. Firstly, the mass term of the `free' fermionic field is
not a constant, thus the inhomogeneity of space already changes the
main assumptions, like power counting behaviour. Note that in our case
there is no unique large momentum behaviour for the propagator.  And
secondly, the would be external field ${\tilde F}$ appears (after
diagonalization) in such a way that it plays a similar role to the
inhomogeneous mass. Then the external field is more likely to appear
in a similar way to a mass in a standard divergent expression, which
is indeed a non-polynomial dependence.
In spite of this, a renormalization prescription may indeed be used for this 
quantity. For example, we may realize that $W_\epsilon$, if regarded as a function of 
$E_2$, has divergences where $E_2$ appear as a pole, a constant or a linear term
in $E_2$. Thus we may cancel all the divergences of this system by including
three counterterms, namely, by adding to $W_\epsilon$ the 1-loop counterterm action
$\delta W_\epsilon$ defined by:
\begin{equation}
  \label{counter}
\delta W_\epsilon \;=\; \alpha^{-1} \, (E_2)^{-1} \,+\, \alpha_0 \,+\, \alpha_1 E_2  
\end{equation}   
where the $\alpha_j$'s are divergent constants. These three constants
require the use of some renormalization conditions to fix them; in our
case one could of course use the Laurent expansion of the actually
measured $W$ in order to fix those coefficients. It is amusing to
note that this procedure requires the knowledge of the full dependence
of $W_\epsilon$ on $\tilde{F}$, since the divergent part is not just a single
polynomial in $\tilde{F}$.

\section{The general case}\label{sec:general}
Let us call $\Fd_\|$ and $\Fd_\bot$ respectively the projections of $\Fd$
onto the direction of $\n$ and onto a direction orthogonal to $\n$
(with respect to the scalar product defined in
section~\ref{sec:first}) Then the square modulus of the Dirac operator
for general configuration of the external vectors, can be written in
the form,

\begin{equation}\label{Hgral}
{\mathcal
H}=-\partial^2+\frac{1}{4}\{x^2(\Fd_\bot^2+\Fd_\|^2)-[x.(\Fd_\bot+\Fd_\|)]^2\}+
(x.\n)^2-ix \times \partial.(\Fd_\bot+\Fd_\|)+\gamma.(\Fd-\n).
\end{equation}

In the same way as before we define versors $\hat{\n}$, $\hat{t}$ and
$\hat{b}$, where now $\hat{t}_{\mu}$ is a unit vector in the direction of
$\Fd_\bot$, i.\ e.\ in the direction

\begin{equation}
\Fd_\mu - \Fd \cdot \; \hat{\n}\; \hat{\n}_\mu.
\end{equation}

We can now define creation and destruction operators $a_{\alpha}$ and
$a_{\alpha}^\dagger$, analogously as it has been done in Eq. (\ref{a}), with
$\alpha=\n,t,b$, and $\w_\alpha$ defined by

\begin{equation}\label{wgral}
\w_\n^2=\frac{\Fd_\bot^2}{4}+\n^2,\;\;\;\;\;\;\;
\w_t^2=\frac{\Fd_\|^2}{4},\;\;\;\;\;\;\;\w_b^2=\frac{\Fd_\bot^2+\Fd_\|^2}{4}.
\end{equation}
In terms of these, $\mathcal {H}$ can be written as,

\begin{eqnarray}\label{wgral}
{\mathcal H}=& & \sum_{\alpha}\w_\alpha (2a_\alpha^\dagger a_\alpha + 1) -
4(\frac{\w_t(\w_b^2-\w_t^2)}{\w_n})^{1/2} (a_ta_\n+a_t
a_\n^\dagger+a_t^\dagger a_\n+a_t^\dagger a_\n^\dagger)
\nonumber \\ &
&-i(\w_b^2-\w_t^2)^{1/2}(\sqrt{\frac{\w_\n}{\w_b}}-\sqrt{\frac{\w_b}{\w_\n}})
(a_ba_\n - a_b^\dagger a_\n^\dagger)  \nonumber \\ & &  -
i(\w_b^2-\w_t^2)^{1/2}(\sqrt{\frac{\w_\n}{\w_b}}+\sqrt{\frac{\w_b}{\w_\n}})
(a_b^\dagger a_\n - a_b a_\n^\dagger)  \nonumber \\ & & - i \w_t
(\sqrt{\frac{\w_b}{\w_t}}-\sqrt{\frac{\w_t}{\w_b}}) (a_ta_b -
a_t^\dagger a_b^\dagger) - i \w_t
(\sqrt{\frac{\w_b}{\w_t}}+\sqrt{\frac{\w_t}{\w_b}}) (a_t^\dagger
a_b - a_t a_b^\dagger).\nonumber\\
\end{eqnarray}

We see that ${\mathcal H}$ is again a bilinear form in the creation and 
destruction operators $a_{\alpha}$ and $a_{\alpha}^\dagger$:
\begin{equation}\label{bilinear1}
H\;=\;a_j^\dagger A_{jl} a_l - \frac{i}{2} a_j B_{jl} a_l + \frac{i}{2} a_j^\dagger B^*_{jl} a_l^\dagger + {\rm constant}\;,
\end{equation}
where $A_{jl}$ and $B_{jl}$ are components of $3\times3$ matrices, $A$ Hermitian and $B$ 
symmetric:
\begin{equation}\label{matrices1}
A= \left(\begin{array}{ccc}
  2 \omega_\eta & -t & i v \\
  -t  & 2 \omega_t & -i r\\
   -i v& i r & 2 \omega_b  
\end{array} \right) \;\; 
B= \left(\begin{array}{ccc}
  0 & -i t & u \\
  -i t & 0 & s \\
     u & s & 0
\end{array} \right)\; ,
\end{equation}
with the constants defined by
$$
r \,=\, ( \sqrt{\frac{\omega_b}{\omega_t}} + \sqrt{\frac{\omega_t}{\omega_b}}) \omega_t \;\; 
s \,=\, ( \sqrt{\frac{\omega_b}{\omega_t}} - \sqrt{\frac{\omega_t}{\omega_b}}) \omega_t \;\;
t \,=\, 4 \frac{\omega_t}{\omega_\eta} \, \sqrt{\omega^2_b - \omega^2_t}
$$
\begin{equation}
u \,=\, (\sqrt{\frac{\omega_\eta}{\omega_b}} - \sqrt{\frac{\omega_b}{\omega_\eta}}) \sqrt{\omega^2_b - \omega^2_t}\;\;
v \,=\, (\sqrt{\frac{\omega_\eta}{\omega_b}} + \sqrt{\frac{\omega_b}{\omega_\eta}}) \sqrt{\omega^2_b - \omega^2_t}\;.
\end{equation}
We also see that in this case it is possible to use a transformation,
defined in terms of $3\times3$ matrices $U$ and $V$, such that the Hamiltonian
is diagonalized. The properties of this transformation can be summarized by
the equations: 
\begin{equation}
U^\dagger U - V^\dagger V\,=\,I \;\;\;\; U^t V-V^t U\,=\,0 
\end{equation}
and 
\begin{equation}
A U + i B^* V \,=\, U E \;\;\;\; -i B U + A^* V \;=\;- V E \;\;.
\end{equation}

As in the simpler $2\times2$ case, one may obtain linear equations that determine 
the transformation matrices $U$ and $V$. In order for those equations to have 
a non-trivial solution, it is necessary to demand the condition:
\begin{equation}
  \label{secular3}
\det\left[A^* (B^*)^{-1} A - B + ( (B^*)^{-1} A - A^* (B^*)^{-1}) {\mathcal E} - (B^*)^{-1} 
{\mathcal E}^2 \right] = 0
\end{equation}
where ${\mathcal E}$ are the `energies', i.e., the numbers appearing in the
diagonal of the matrix $E$.
Equation (\ref{secular3}) may be explicitly solved, what yields $6$ solutions.
Of these $6$ solutions, we should eliminate $3$ spurious ones, since they are
not consistent with the algebraic relations that are verified by $U$ and $V$.
This leads to the following three solutions. One of them vanishes: 
${\mathcal E}_1=0$, and the other two have a rather cumbersome expression, which
may be written in terms of the previously defined $\omega$ parameters  as follows:
$$
{\mathcal E}_2 = \sqrt{2}\left\{ \omega_\eta^2 + \omega_t^2 + 3 \omega_b^2 - \; ( \omega_\eta^{-1} [(\omega_b^2-\omega_t^2)( 64 \omega_t^3 + 96 \omega_t^2 \sqrt{\omega_\eta \omega_t})]
\right.
$$
\begin{equation}
\left.
+ \omega_\eta^4 - 14 \omega_\eta^2 \omega_t^2 + 13 \omega_t^4 + 6 (\omega_\eta^2-\omega_t^2)\omega_b^2 + 9 \omega_b^4)])^{\frac{1}{2}} \right\}^{\frac{1}{2}}
\end{equation}
$$
{\mathcal E}_3 = \sqrt{2}\left\{ \omega_\eta^2 + \omega_t^2 + 3 \omega_b^2 + \; ( \omega_\eta^{-1} [(\omega_b^2-\omega_t^2)( 64 \omega_t^3 + 96 \omega_t^2 \sqrt{\omega_\eta \omega_t})]
\right.
$$
\begin{equation}
\left.
+ \omega_\eta^4 - 14 \omega_\eta^2 \omega_t^2 + 13 \omega_t^4 + 6 (\omega_\eta^2-\omega_t^2)\omega_b^2 + 9 \omega_b^4)])^{\frac{1}{2}} \right\}^{\frac{1}{2}} \;.  
\end{equation}

>From these solutions the Hamiltonian operator can be written in
diagonal form and the determinant of the Dirac operator can be
obtained for general $\Fd$ and~$\n$. Calculations in this case are
more involved, but following along the same lines as we have done in
section~\ref{sec:vacuum}, it should be possible to obtain the vacuum
energy in the general case. Progress on this subject will be reported
elsewhere.

\section*{Acknowledgments}
This work was part of a collaboration supported by CLAF (Centro
Latinoamericano de F{\'\i}sica). We also acknowledge financial support by
Fundaci{\'o}n Antorchas, Argentina.

\end{document}